%% file: article.tex
\documentclass[pre, reprint, superscriptaddress]{revtex4-2}

\input{pack_def}

\graphicspath{{figs/}}
\begin{document}

\title{Black to white transition of a charged black hole}
\author{Antoine Rignon-Bret}
\affiliation{\'{E}cole Normale Sup\'{e}rieure, 45 rue d'Ulm, F-75230 Paris, France}
\author{Carlo Rovelli}
\affiliation{Aix-Marseille University, Universit\'e de Toulon, CPT-CNRS, F-13288 Marseille, France.}
\affiliation{Department of Philosophy and the Rotman Institute of Philosophy, 1151 Richmond St.~N London  N6A5B7, Canada}
\affiliation{Perimeter Institute, 31 Caroline Street N, Waterloo ON, N2L2Y5, Canada}

\begin{abstract}\noindent
We present an exact solution of the Maxwell-Einstein equations, which describes the exterior of a charged spherical mass collapsing into its own trapping horizon and then bouncing back from an anti-trapping horizon at the same space location of the same asymptotic region.  The solution is locally but not globally isometric to the maximally extended Reissner-Nordstr\"{o}m metric and depends on seven parameters. It is regular, and defined  everywhere except for a small region, where quantum tunnelling is expected. This region lies \emph{outside} the mass: the mass-bounce and its near exterior are governed by  classical general relativity.  We discuss the relevance of this result for the fate of realistic black holes. We comment on possible effects of the classical instabilities and the Hawking radiation. 

\end{abstract}

\maketitle

\section*{Introduction}
\label{intro}

Contrary to what is sometimes assumed, the long term evolution of the \emph{exterior} of a black hole is likely to be affected by quantum gravitational effects.  This is because the back-reaction of the Hawking radiation makes the curvature grow to Planckian values \emph{outside} the horizon too. At this point, and its immediate future, the Einstein equations are likely to be violated by quantum gravity effects, in particular by conventional quantum tunnelling. Elsewhere, it is reasonable to expect the evolution of spacetime to be governed by the \textit{classical} Einstein equations. What  do \emph{these} permit for the long term evolution of a black hole?  

A surprisingly result obtained in \cite{Haggard2014} is that the  Einstein equations admit a spherically symmetric vacuum solution that describes a black hole that tunnels into a white hole.  This spacetime does not contradict Birkhoff's theorem, because it is locally --but not globally-- isometric to the Kruskal spacetime.  The result reveals an interesting scenario: (i) black hole horizons are \emph{event} horizons  only in the classical limit; (ii) before the end of the evaporation, the trapping horizon tunnels into an antitrapping horizon;  (iii) the black hole interior tunnels into a white hole interior; (iv) the collapsed matter bounces out. This scenario \cite{
Rovelli2014ps, christodoulou2016realistic, Rovelli2018h, Bianchi2018e, Christodoulou2018d, Rovelli2018a, Martin-Dussaud2019, Dambrosio2021, soltani2021end}, its possible astrophysical implications  
\cite{Barrau2014b,Barrau2014c,Vidotto2016,Rovelli2017f,Barrau2017,Raccanelli2017,Vidotto2018a,Rovelli2018f,Rovelli2018g,Vidotto2018b,Barausse2020,barrau2021closer}, and its relevance for the black hole information paradox  \cite{Rovelli2017a,Rovelli2019a} are currently under intense investigation.   

So far the literature has focused on the non-rotating and non-charged case. Realistic black holes rotate and are approximated by the Kerr-Newman metric, whose maximal extension has a markedly different structure than the Kruskal spacetime.  Is the intriguing black to white transition a peculiarity of the Schwarzschild metric, or is it possible in general? 

The lack of spherical symmetry makes the analysis of the Kerr-Newman case harder, but there is an interesting intermediate case that has the same global structure as extended Kerr (the Carter-Penrose diagrams of their maximal extension is similar) and yet has spherical symmetry: the Reissner-Nordstr\"{o}m metric. This is the solution of the Maxwell-Einstein equations around a spherical symmetric \emph{charged} mass.  In this paper, we extend the result of \cite{Haggard2014} to this case. We show that there is an exact solution of the Einstein-Maxwell equations that  describes the exterior of a charged spherical mass that collapses into its own trapping horizon and then bounces back from an anti-trapping horizon at the same spacial location of the same asymptotic region.  The solution is locally but not globally isometric to the maximally extended Reissner-Nordstr\"{o}m metric, is regular, and is everywhere defined except for a compact finite region, where a quantum gravitational tunnelling transition can be expected.

What we find is surprising.   Unlike the non-charged case, the quantum region does not continue inside the hole all the way to the collapsing and bouncing matter.  The bounce of the collapsing matter and its surrounding evolve \textit{classically} without ever entering the quantum region.   This is comprehensible, as the global structure of Reissner-Nordstr\"{o}m (and Kerr) allows a time-like geodesic to enter a black hole, traverse it and exit from a white hole without  encountering singularities or high curvature regions. In the maximally extended metric, the white hole is in a different asymptotic region; here we show (following  \cite{Haggard2014}) that the white hole can be in the same asymptotic region as the black hole: in its immediate future.

This result makes the charged case transition (and presumably the rotating case as well) easier to understand and treat than the Schwarzschild case. In a sense, the spacetime region which needs to be described by quantum gravity is smaller than in the Schwarzschild case: some part of the mechanism of the black to white transition is already contained in the classical solution. Quantum effects are not needed for the charged mass to bounce, nor for the black hole interior to evolve into a white hole exterior.  Only the horizon area undergoes a quantum transition, when it reaches  Planckian curvature. 

We recall the main features of the black to white transition in the Schwarzschild case in Section~\ref{Schbw} and the causal structure of the maximally extended Reissner-Nordstr\"{o}m metric in Section~\ref{The maximally extended Reissner}.  The solution of the Maxwell-Einstein equations that describes the black to white transition of the charged black hole is built in   Section~\ref{buildmetric} and the reason a quantum tunnelling is to be expected is discussed in Section~\ref{PP}. Then we comment the possible effects of the classical instabilities  \cite{penrose1999question,poisson1989inner, poisson1990internal} and the Hawking evaporation in Sections~\ref{insta} and ~\ref{HR}. We use units where $G$ = $c$ = $\hbar$ = 1.

\section{Schwarzschild}
\label{Schbw}

 \begin{figure}
    \centering
    \includegraphics[width=3cm]{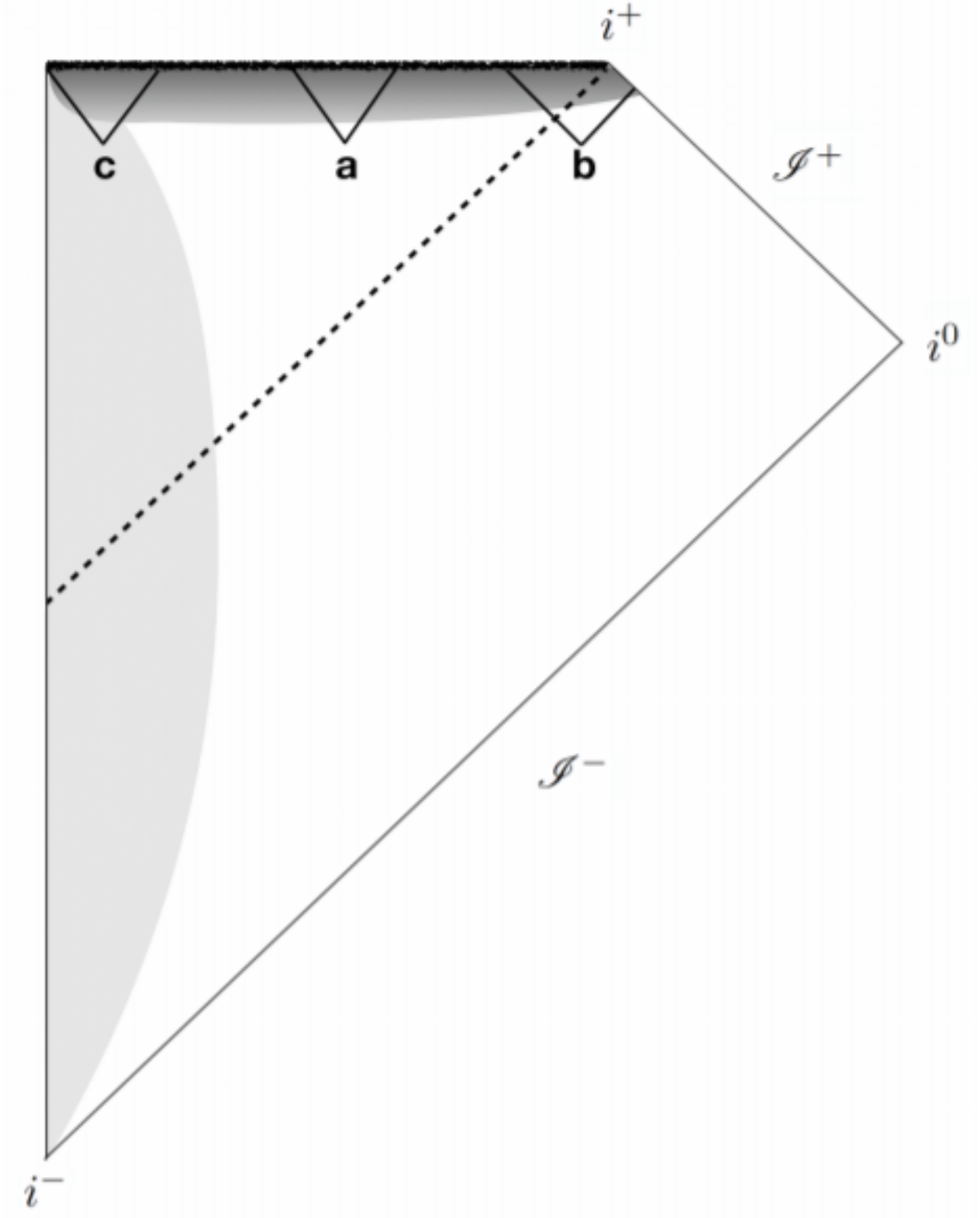}
    \caption{The Carter-Penrose diagram of a Schwarzshild black hole until the onset of quantum gravity. The light grey region is the collapsing star. The dark grey region is where quantum gravity becomes relevant.}
    \label{SchBH}
\end{figure}

The Carter-Penrose diagram of a (classical) Schwarzschild black hole created from a gravitational collapse is depicted on Figure~\ref{SchBH}. The dark grey region is where the classical theory becomes unreliable, due to quantum gravitational effects. We expect this to happen when the curvature becomes Planckian, for instance when the Kretschmann scalar 
\begin{equation}
    K^2 = R_{\alpha \beta \gamma \delta}R^{\alpha \beta \gamma \delta} = 48\frac{M^2}{r^6}
    \label{kretscalar}
\end{equation}
becomes of order 1. Here  $M$ is the black hole mass and $r$ is the Schwarzschild radius.  This happens before the $r=0$ singularity inside the black hole. Just outside the horizon, $K \sim \frac{1}{M^2}$. The Hawking evaporation steadily decreases the mass $M$ of an isolated black hole, bringing it down to Planckian values, hence the quantum region extends  outside the horizon. 
General arguments and some specific calculation \cite{Bianchi2018e, christodoulou2018characteristic} indicate that the transition probability $P$ from black holes to white holes is proportional to 
\begin{equation}
    P \sim e^{-M^2}.
    \label{probatuunel}
\end{equation}
Thus becoming dominant at the end of the evaporation, where $M\sim 1$.  It is also possible that quantum effects could appear earlier  \cite{Haggard2014, Haggard2016}, at a time of order $M^2$ after the collapse.   

\begin{figure}
    \centering
    \includegraphics[width=3cm]{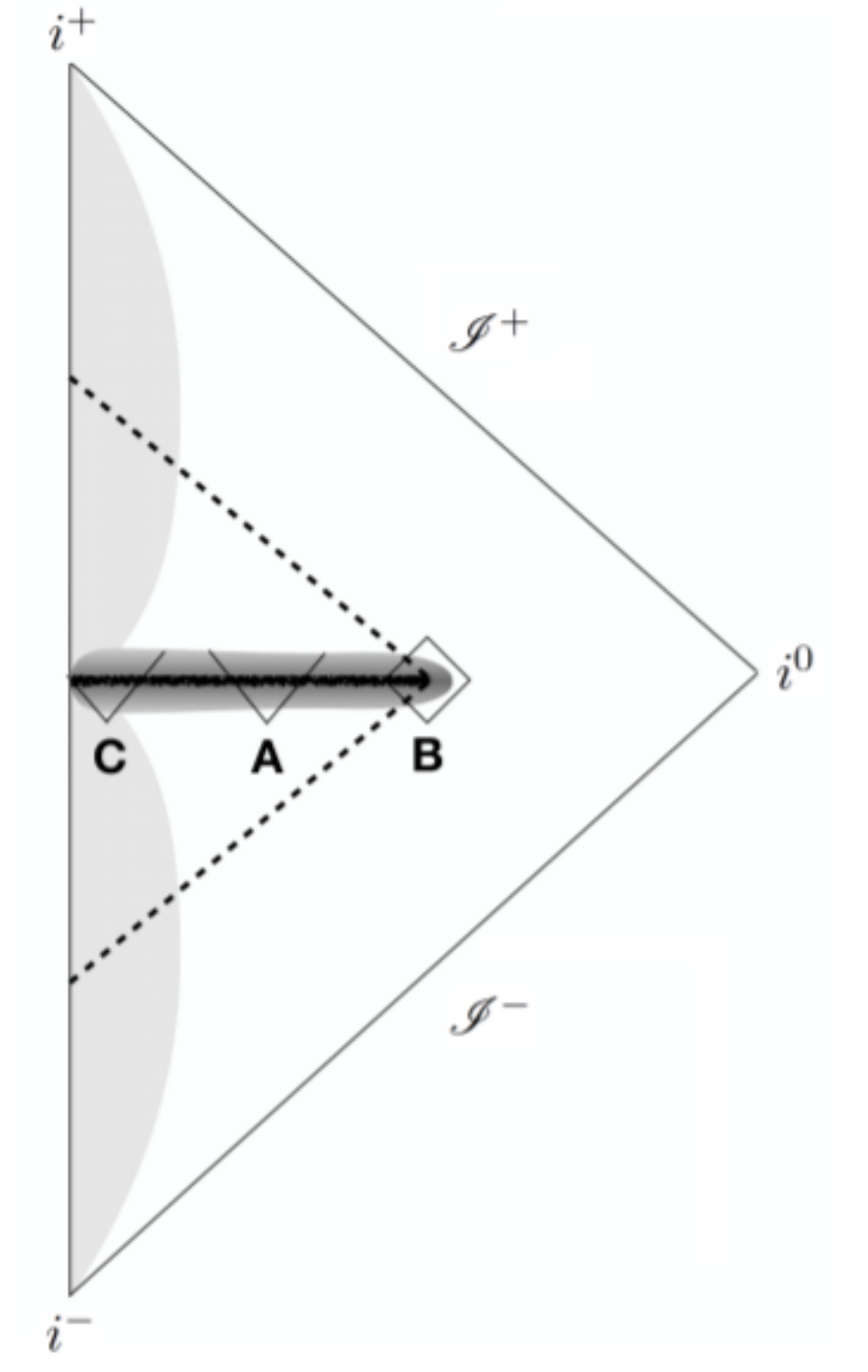}
    \caption{The Carter-Penrose diagram of the black to white transition. The dark grey region is the quantum gravity region. The black hole (trapped region) is below the quantum gravity region while the white hole is above. The trapping horizons are the dashes lines.}
    \label{b2wSch}
\end{figure}

Here we are not directly concerned with these estimates. We only retain the fact that the quantum region extends outside the horizon. 
This region can be organized into three sub regions \cite{Dambrosio2021} (see Figures \ref{SchBH} and  \ref{b2wSch}):
\begin{itemize}
\setlength\itemsep{.5mm}
\item Region \textbf{B} : The horizon region. 
  \item Region \textbf{C} : The  collapsing star region. 
  \item Region \textbf{A} :  The region which is neither directly causally connected to the horizon nor to the collapsing star.
  \end{itemize}
It is shown in \cite{Dambrosio2021} that the phenomena in these three regions can be considered  causally disconnected, as they are separated by a large spacelike distance, which at the end of the evaporation grows to 
\begin{equation}
    L \simeq  {M}^\frac{10}{3},
    \label{interior}
\end{equation}
which is huge for a macroscopic black hole.  To understand the physics of the end of a black hole, we have to understand the quantum evolution of these three regions of spacetime. 

The Carter-Penrose diagram of the classical metric describing the black to white bounce \cite{Haggard2014} is depicted in Figure \ref{b2wSch}.  The white region is locally but not globally isomorphic to the Kruskal metric, the maximal extension of the Schwarzshild metric. There is a trapped as well as a later anti-trapped region  (see also Figure 8 in \cite{Carballo-Rubio2020}), separated by a small compact region where the Einstein equations are violated by quantum gravity effects. The horizon of the black hole is not an event horizon. 

\section{The maximally extended Reissner-Nordstr\"{o}m metric}
\label{The maximally extended Reissner}

\begin{figure}
    \centering
    \includegraphics[width=0.4\linewidth]{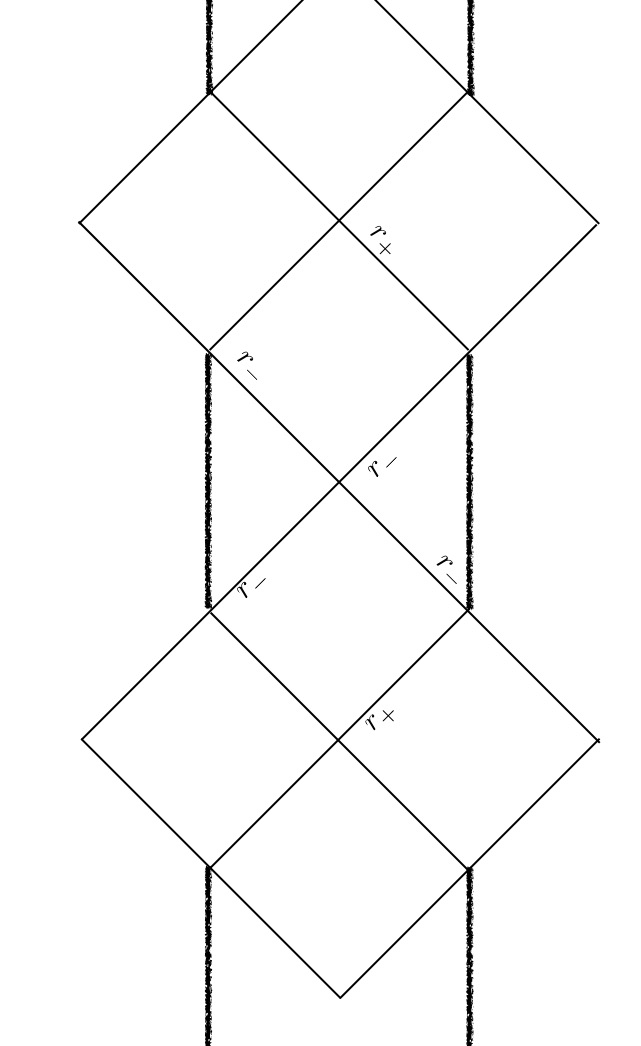}
    \caption{Maximally extended Reissner-Nordstr\"{o}m metric}
    \label{NR1}
\end{figure}

The Reissner-Nordstr\"{o}m metric describing a massive charged star (or black hole) is  
\begin{equation}
    ds^2 = -f(r) dt^2 + \frac{1}{f(r)} dr^2 + r^2 d\Omega^2
    \label{reissnermetric}
\end{equation}
with 
\begin{equation}
    f(r) = 1 - \frac{2M}{r} + \frac{Q^2}{r^2},
    \label{reissner term}
\end{equation}
where $M$ is the mass of the star and $Q$ its charge. Together with the Maxwell potential 
\begin{equation}
    A_a = \left(\frac{Q}{r},0,0,0\right),
    \label{A}
\end{equation}
equations (\ref{reissnermetric}--\ref{reissner term}) solve the Maxwell-Einstein equations.  

Notice that the $Q\to 0$ limit that reduces Reissner-Nordstr\"{o}m to Schwarzschild is subtle at small radius, as 
 \begin{equation}
  \lim_{Q\to 0}  \lim_{r\to 0}f(r)= \infty,\ \ \  {\rm while} \ \ \  
  \lim_{r\to 0}  \lim_{Q\to 0}f(r)= -\infty,
     \label{lim2}
\end{equation}
 which shows that even a small charge changes the inner geometry radically. 

As Schwarzschild, the Reissner-Nordstr\"{o}m  metric is static outside the horizon, and spherically symmetric. The main difference with Schwarzschild is that the equation $f(r) = 0$ that gives the position of the trapping horizons has two solutions rather than one: 
\begin{equation}
    r_{\pm} = M \pm \sqrt{M^2 - Q^2}.
    \label{outer/inner horizon}
\end{equation}
The two solutions $r_+$ and $r_-$ define the outer and inner horizon, respectively. They separate  trapped, non-trapped and anti-trapped regions: the surfaces of constant $r$-coordinate are timelike for $r > r_+$, becomes spacelike for $r_- < r < r_+$, and are timelike again for $r < r_-$. 
 The difference with Schwarzschild is therefore the presence of the inner horizon $r_-$, a feature in common with the  Kerr metric.  As for Schwarzschild, the metric has four Killing fields, the three  associated to the rotations, and the Killing field $\frac{\partial}{\partial t} = \xi$ associated to the invariance with respect to the $t$ coordinate. Notice that $\xi$ is timelike for $r > r_+$ and $r < r_-$, null at the two horizons and spacelike for $r_- < r < r_+$.
 
The Penrose diagram of the Maximally extended Reissner-Nordstr\"{o}m spacetime is given in Figure~\ref{NR1}. The interior region continues into an anti-trapped region, namely a white hole region, that in turns exits via another outer horizon into a different asymptotic region. 
 
\begin{figure}
    \centering
    \includegraphics[width=0.4\linewidth]{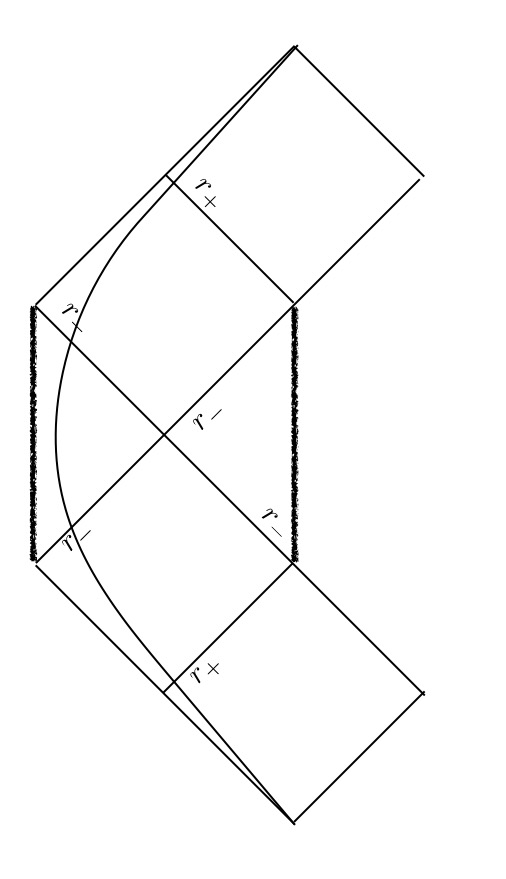}
    \caption{The portion of the maximally extended Reissner-Nordstrom spacetime relative for the black to white bounce. The grey line is a geodesic that enters the black hole from a asymptotic region, and exit in a different asymptotic region.}
    \label{NR2}
\end{figure}

Consider a neutral particle with mass  $m$ and energy momentum $p_\alpha= m u_\alpha$ falling into a Reissner-Nordstr\"{o}m black hole. The quantity $E = - g_{\alpha \beta} p^\alpha \xi^{\beta}$ is a constant of motion.  A straightforward calculation gives 
\begin{equation}
    \dot{r}^2 + f(r) = E^2,
    \label{equationmotion}
\end{equation}
where $ \dot{r}=u^r$.  Notice that $ \dot{r}$ vanishes at the radius $r_b$ determined by 
\begin{equation}
   f(r_b) = E^2.
\end{equation}
If the initial radial velocity of the particle vanishes at large $r$, then $E=1$ and
\begin{equation}
   r_b= \frac{Q^2}{2M}<r_-.
   \label{rb}
\end{equation}
Hence, the particle enters the black hole, crosses the outer horizon at $r_+$, then crosses the inner horizon at $r_-$, and  reaches $r = r_b$ where its velocity goes to zero. The particle bounces and starts moving outward.  This is permitted because the interior region is not trapped. By time inversion symmetry, its geodesics exit then into the next asymptotic region through the anti-trapped region.  See Figure \ref{NR2}.

The conclusion is also valid for a charged particle. Due to \eqref{A}, the electromagnetic effect on the energy of a charged particle is simply to shift it to $p_0 = mu_0 + qA_0$ where $A_0 = \frac{Q}{r}$ (we use units where $\frac{1}{4 \pi \epsilon_0} = 1$). The conserved quantity is $E = - g^{\alpha \beta} p_\alpha \xi_{\beta} = - p_\alpha \xi^{\alpha} $, giving 
\begin{equation}
    \bigg( {E} - \frac{Q^2}{M r} \bigg)^2 = f(r) + \dot{r}^2. 
    \label{motion2}
\end{equation}
Charged particles do not follow geodesics, but \eqref{motion2} is similar to \eqref{equationmotion} up to a shift in the energy. If the initial radial velocity vanishes, it implies from \eqref{motion2} that ${E} \simeq 1$ by assuming that the initial radius is sufficiently large. Hence, we have to solve Eq.~\eqref{motion2} with the turning point condition ($\dot{r}^2 = 0$), which gives 
\begin{equation}
    \bigg( 1 - \frac{Q^2}{M r} \bigg)^2 = f(r)
    \label{motion3}
\end{equation}
Interestingly, the solution of this equation is still~\eqref{equationmotion} where we didn't take the electrostatic repulsion into account.

  \begin{figure}
    \centering
    \includegraphics[width=0.4\linewidth]{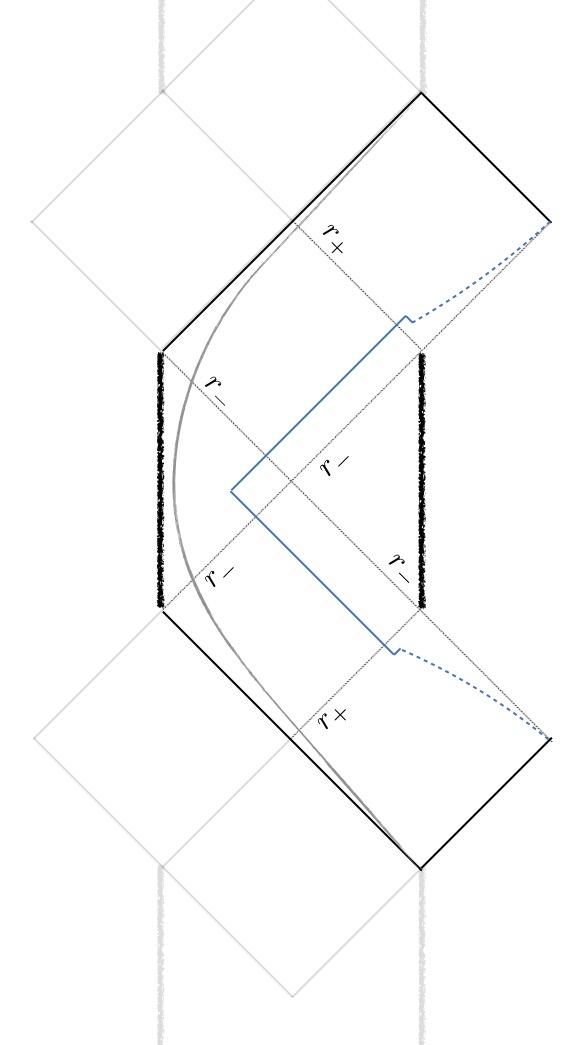}
    \caption{The surface of the star (grey line), the boundary of the quantum region (continuous blue line) and the constant time surface that can be identified.}
    \label{NR6}
\end{figure}

Thus, a collapsing massive spherical charged mass that enters its own outer horizon can cross the inner horizon and  bounce out at $r_b$.

In the Schwarzschild case the bounce of the star was an hypothesis based on quantum gravity.   In the charged case, the bounce of the star is predicted by \emph{classical} general relativity.  The physical problem of the {\bf C} region defined above is solved without the need of quantum mechanics. 

In other words, the presence of a charge opens up a \emph{classical} throat for the passage of the star and its surrounding spacetime from the black to the white hole region.  

If the charge is small this passage is narrow. Its size can be estimated from the spacelike distance of the boundary of the star to the outer boundary of the classical region on a constant $t$ surface. This is 
\begin{eqnarray}
    d &=&  \int_{r_b}^{r_-}  \frac{dr}{\sqrt{f(r)}} =
    \int_{r_b}^{r_-} \frac{r}{\sqrt{(r - r_+)(r - r_-)}} dr \nonumber
    \\ && \leq  \int_{0}^{r_-} \frac{r_-}{\sqrt{(r_- - r_+)(r - r_-)}} dr \nonumber
    \\ &&  = \sqrt{2} M \frac{\bigg(1 - \sqrt{1 - \eta^2} \bigg)^\frac{3}{2}}{(1 - \eta^2)^\frac{1}{4}}
    \label{sizealpha}
\end{eqnarray}
where $\eta = \frac{Q}{M}$. If $\eta$ is small, the throat is small. Only matter falling just after the star collapses reaches it, rather than the Cauchy horizon, But if $\eta$ is close to 1\begin{equation}
    ds = \frac{r}{\sqrt{(r -r_+)(r -r_-)}} dr \simeq \frac{r}{r - r_-} dr
    \label{extremRN}
\end{equation}
which is not integrable in $r_-$. Hence the throat can be arbitrarily large.

\begin{figure}
    \centering
    \includegraphics[width=0.5\linewidth]{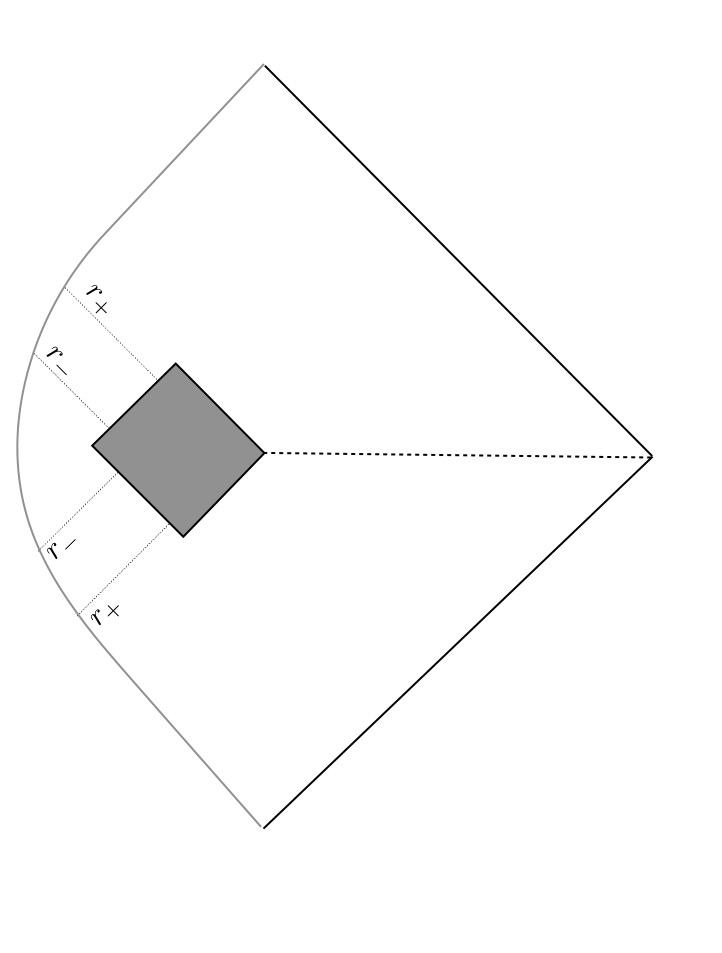}
    \caption{The spacetime of the black to white charged transition outside the star.}
    \label{NR7}
\end{figure}

The result suggests that the Reissner-Nordstr\"{o}m black to white transition may be more natural than the Schwarzschild one: allready in the classical theory the white hole is  in the future of the black hole.

\section{The Reissner-Nordstr\"{o}m  black to white transition}
\label{buildmetric}
\label{compaSchRN}

Here we are not interested in an hypothetical emergence in a different universe.  We are interested in the compatibility of  classical general relativity with a bounce within the \emph{same} universe, generated by a quantum tunnelling  in a small compact spacetime region.

To show that this is possible, we can construct a solution of the Maxwell-Einstein equations, following \cite{Haggard2014},  by cutting and gluing relevant portions of the maximally extended Reissner-Nordstr\"{o}m metrics.  The way this can be done is sketched in Figure \ref{NR6}.

\begin{figure}
    \centering
    \includegraphics[width=0.5\linewidth]{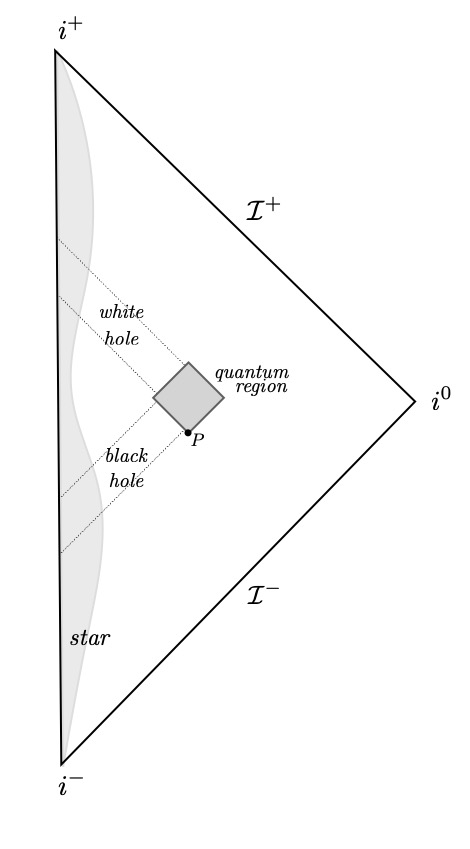}
    \caption{The full black to white transition, for a charged black hole.}
    \label{full}
\end{figure}

To construct the spacetime we are interested in, we proceeded as follows. (i) We cut away from the maximally extended Reissner-Nordstr\"{o}m spacetime all the regions on the left of the grey line in Figure \ref{NR6}, and replace it with the interior of a classical bouncing star.  (ii) We cut away all the region to the right of the blue line in Figure \ref{NR6}. (iii) We glue the two dotted portions of the blue line. These are both constant-Schwarzschild-time surfaces, and therefore the gluing gives a smooth junction, where the Einstein equations are satisfied.  

More precisely, to glue the spacetime smoothly both metric and extrinsic curvature must match. The metric of two surfaces of constant time coordinate $t$ is clearly the same.  By construction, the extrinsic curvature $k$ changes sign (because of the time reversal symmetry exploited in cutting the spacetime). But the curvature vanishes on these surfaces as the normal to the spacelike surfaces of constant $t$ is the vector $N_\alpha = \partial_\alpha t = \xi$ where $\xi$ is the Killing vector associated to the $t$-coordinate invariance and 
\begin{equation}
    k = \frac{1}{2} \mathcal{L}_{N} q= \frac{1}{2} \mathcal{L}_{\xi} q= 0
\end{equation}
where $q$ is the induced metric on the spacelike hypersurface.

The resulting spacetime is depicted in Figure  \ref{NR7}, and in  \ref{full} by including the star.  The central grey area represents the quantum tunnelling region, where the classical evolution is violated.  Let us do so more explicitly.

\begin{figure}
    \centering
    \includegraphics[width=0.8\linewidth]{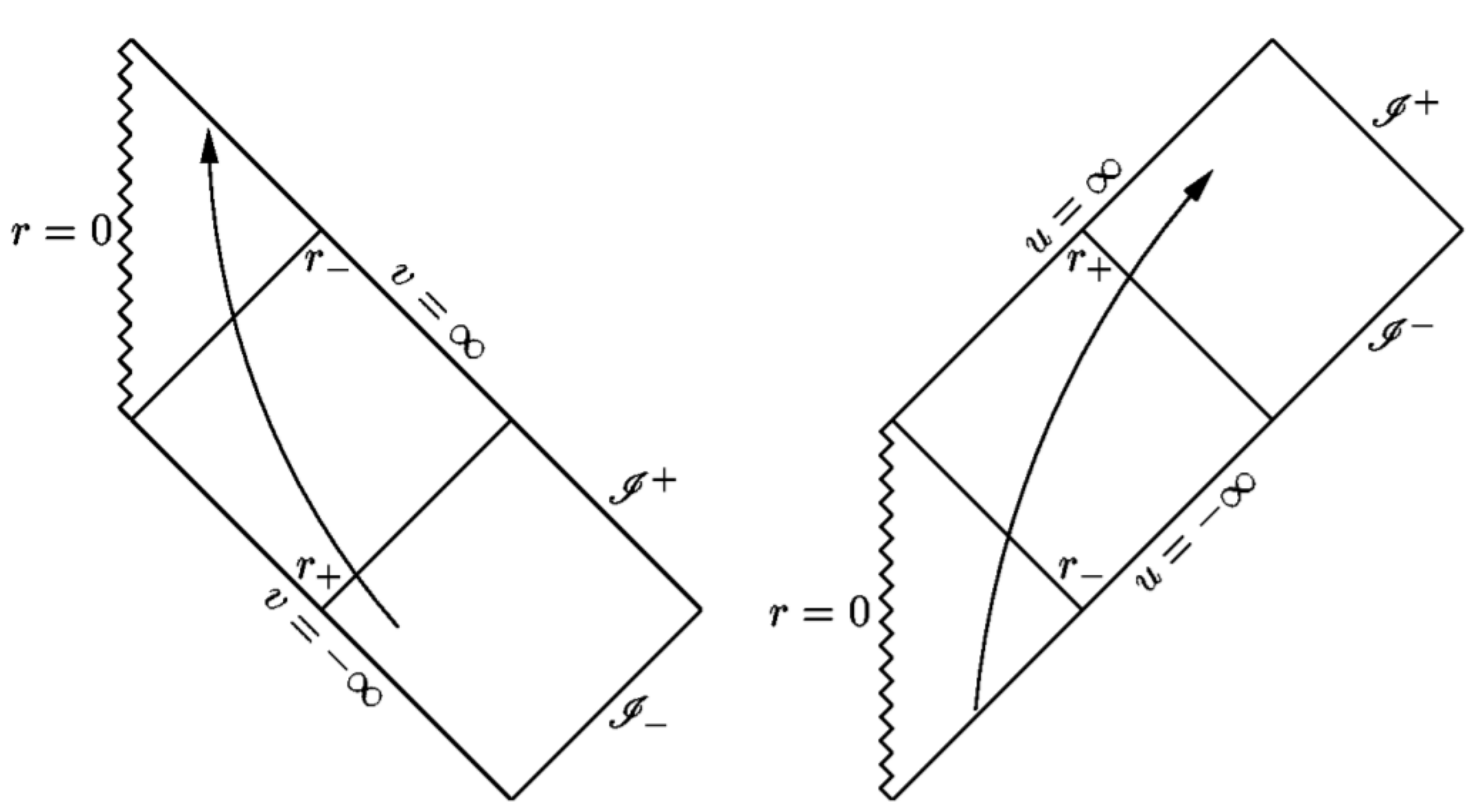}
    \caption{The part of the maximally extended Reissner-Nordstr\"{o}m spacetime surrounding to the black hole region on the left, that can be  labeled by $(r, v)$ coordinates, and the part corresponding to the white hole region on the right,  by $(r,u)$ patch. Picture taken from \cite{poisson2002advanced}.}
    \label{patch}
\end{figure}

Let us break the spacetime of Figure \ref{NR6} into two overlapping regions, as depicted in Figure \ref{patch}, and introduce distinct coordinates in the two regions.  

The left panel includes the exterior  $r>r_+$, the black hole $r_-<r<r_+$,  and the interior $r<r_-$.  A set of coordinates covering this entire left region is given by $r$ and the advanced time null coordinate $v$ defined by
\begin{equation}
    v = t + r^*(r)
    \label{v}
\end{equation}
where 
\begin{equation}
    r^*(r) = \int \frac{dr}{f(r)}.
    \label{tortoisec}
\end{equation}
The integration must be done separately in the regions separated by $r_+$ and $r_-$, because the integral is ill behaved at the horizons.   (Remember that the $t$ coordinates in the three regions are actually unrelated to one another, each going from $-\infty$ to $+\infty$ in its own region. Properly speaking, they should have different names.)  In the interior region, it is convenient to choose the integration constant (or the lowest boundary of the integral) so that the $t$ coordinate of the bounce point $b$ is zero.  In the external region, it is convenient to choose it so that the $t$ coordinate of the outermost point of the quantum region is zero.

 In these  coordinates the metric reads 
\begin{equation}
    ds^2 = - f(r) dv^2 + 2 dr dv + r^2 d\Omega^2. 
    \label{rvmetric}
\end{equation}

The right panel of Figure \ref{patch} includes the exterior, the white hole and again the interior region, separated by $r_+$ and $r_-$.   A set of coordinates covering the left region is given by $r$ and the retarded time 
\begin{equation}
    u = t - r^*(r).
    \label{u}
\end{equation}
In these coordinates, the metric reads 
\begin{equation}
    ds^2 = - f(r) du^2 - 2 dr du + r^2 d\Omega^2.
    \label{rumetric}
\end{equation}
The two coordinate patches overlap in the interior ($r<r_-$)  and exterior ($r>r_+$) regions, where their relation is easily deduced by equating the $t$ coordinate of the two:
\begin{equation}
  u  +  r^*(r)  =  v - r^*(r). 
    \label{t}
\end{equation}
Notice that identifying the coordinates in the two overlapping regions (exterior and interior) adds a parameter to the definition of the spacetime (irrespectively from the detailed location of the quantum region).  This can be seen as follows.  Consider two points $p_1$ and $p_2$ both on $t=0$ (hence $u_1=-v_1$ and $u_2=-v_2$), and let $r_1$ and $r_2$ be their radius.   If the two points are both in the exterior or both in the interior region, the difference of their retarded (or advanced) time is 
\begin{equation}
u_2-u_1=    r^*(r_2)-r^*(r_1) = \int_{r_1}^{r_2} \frac{dr}{f(r)}.
\end{equation}
But if $p_2$ is in the interior and $p_1$ in the exterior, the above integral must be broken into two and therefore depends on an arbitrary integration constant, not determined by the radius of the points.  This additional parameter is a global topological parameter which distinguished spacetimes with the same $M$ and $Q$, obtained in this manner.  

\begin{figure}[t]
    \centering
    \includegraphics[width=0.6\linewidth]{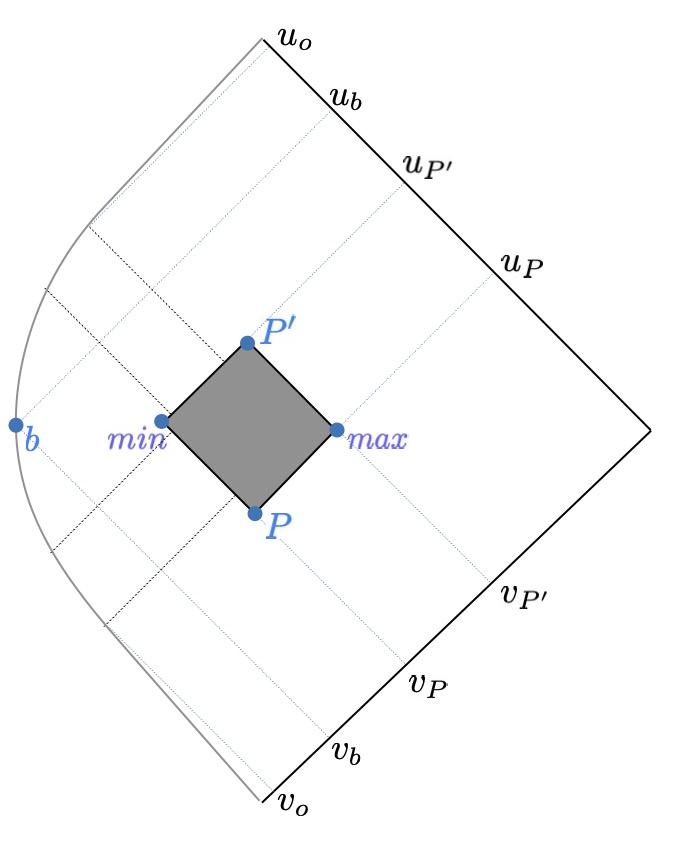}
    \caption{The parameters characterising the bouncing geometry.}
    \label{RN8}
\end{figure}

To specify the metric  entirely (including the location of the quantum region) is it convenient to focus on three special events: the event $b$ where the radius of the star reaches its mininal value $r_b$ given in \eqref{rb}, the event $P$ where the quantum tunnelling starts and the point $P'$ where it ends. We take the quantum region to be the causal diamond defined by $P$ and $P'$, namely the intersection between the causal future of $P$ and the causal past of $P'$: see Figure \ref{RN8}.

Let the coordinates of the points $b,P$ and $P'$ be $(v_b,u_b),(v_P,u_P)$ and $(v_{P'},u_{P'})$, respectively.  The two points of the quantum region $min$ and $max$ with the minimal and maximal radius have coordinates $(v_P,u_{P'})$ and  $(v_{P'},u_P)$ respectively. 

As a first step, consider a time reversal symmetric bounce geometry. In this case we can assume $u_b=-u_v, u_{P'}=-v_P$ and $u_P=-v_{P'}$ (see Figure \ref{RN8}). The maximal radius reached by the quantum region is $r_{max}=(v_{P'}-u_P)/2>r_+$, at $max$. The minimal radius reached by the quantum region is $r_{min}=v_{P}-u_{P'}<r_-$, at $min$.   We can chose the zero of the Schwarzschild time coordinate $t$ so that $b$, $min$ and $max$ are all on $t=0$.

Let us discuss the interpretation of these quantities. The quantities $v_0$ and $u_0$ are the advanced time of the collapse of the star into the black hole and the retarded time of exit from the white hole. 
The difference $v_b-v_o$ is the (short) time determined by the details of the dynamics of the collapsing star. For simplicity, we can take the limit of a star collapsing at very high relativistic speed so that $v_b- v_o\sim 0$ and  neglect their difference.  

The advanced time $\tau_t=v_{P'}-v_P$ is the time of the tunnelling transition (likely to be short).
The crucial physical parameter of the process is the difference $\tau_{BH}=v_P-v_b$.   Notice that we must have $r_{min}<r_-$, but as $r_{min}$ approaches $r_-$, the advanced time $v_P$ goes to infinity.  Hence $\tau_{BH}$ can be arbitrary long.

Disregarding the time  $v_b-v_o$, the symmetric bounce is then fully characterized by five parameters: $M,Q, r_P, \tau_{BH}$ and $\tau_t$.  A theory of quantum gravity must give the transition probability 
$W(M, Q, r_P,\tau_{BH},\tau_t)$ between the black hole state and the white hole state.  For the formulation of this computation in Loop Quantum Gravity, see   \cite{Bianchi2018e, 
Christodoulou2018d, christodoulou2018characteristic, soltani2021end} 

A non-time-reversal-symmetric bounce can be obtained by having $b$ and $(v_P,u_{P'})$ on different constant $t$ surfaces and $P$ and $P'$ at different radius.   This gives $u_{P'}\ne -v_P$ and $\tau_{WH}=u_b-u_{P'}$ different from $\tau_{BH}$.  The two long times $\tau_{BH}$ and $\tau_{WH}$ can be interpreted as the lifetimes of the black hole and the white hole.  The general case is therefore characterized by seven parameters:  $M,Q, r_P,r_{P'},\tau_{BH}$ and $\tau_{WH}$, plus the tunnelling time $\tau_t$.

\section{The point $P$ and the onset of the tunnelling}
\label{PP}

The spacetime described above is characterized by the region $(u_{P}<u<u_{P'},v_P<v<v_{P'})$ where the Einstein equations are not satisfied. Following \cite{soltani2021end}, we call this region the {\bf B} region. As mentioned in the introduction, this is the region where we assume quantum mechanics alters the classical spacetime dynamics.  Why quantum theory should be relevant here? There are three answers to this question:
\begin{enumerate}
\setlength\itemsep{.2mm}
  \item[(i)] Quantum mechanics allows tunnelling transition with a given finite probability. A rough estimate of a the tunnelling probability may include a factor $e^{-S}$, where $S$ is the action of the process. Since $Q$ is of little relevance near the outer horizon, we may expect this factor to be proportional to $e^{-M^2}$.  In the classical theory, the lifetime of a black hole is infinite and during an infinite time, an event may happen even if its probability per unit of time is very small.  This would give an exponentially large factor $e^{M^2}$ in the black hole lifetime, but still a finite lifetime. 
  \item[(ii)]  However, we expect an isolated black hole to evaporate by Hawking radiation. During the evaporation, the mass of the hole decreases, reaching a Planckian value in a time of the order $M_0^3$, where $M_0$ is the initial mass of the hole. When $M$ approaches the Planck mass, the suppressing factor $e^{-M^2}$ approaches unit, and the tunnelling becomes likely to happen. Hence we may expect a transition to happen shortly before the end of the Hawking evaporation, hence in a time of order $M_0^3$ after the collapse.  It is important to recall again that when $M$ is near the Planck mass, the curvature becomes Planckian \emph{outside} the horizon. We can view the point $P$ as a point where the curvature is close enough to a Planckian curvature to trigger the quantum tunnelling. 
  \item[(iii)] Finally, we also recall the hypothesis considered in \cite{Haggard2014}: quantum corrections to the Einstein equations can be small but are never exactly zero, so they can pile up in proper time $\tau$.  Assuming a linear deviation to first order in $\hbar$, a non negligible quantum phenomenon may start when $q=K \tau$ is Planckian, where $K$ is a curvature scalar such as the 
Kretchman invariant 
\begin{equation}
    K^2 = C_{\alpha \beta \gamma \delta}C^{\alpha \beta \gamma \delta} = 48\frac{\left(M -\frac{Q^2}{r}\right)^2}{r^6}
    \label{Kritchmaninvar}
\end{equation}
and $\tau$ is the proper time along a stationary time-like geodesic. Since the curvature near the horizon is of order $M^{-2}$, this has suggested the (speculative) possibility of a transition triggered already after a black hole lifetime of order $M_0^{2}$.    This consideration suggests also what is the possible radius of $P$.  The parameter $q$ is maximized at a finite distance from the horizon, of the order of $M$.  This is because the curvature increases with smaller radius, but the proper time at a stationary point is red shifted near the outer horizon as $\tau=\sqrt{f(r)}\ \tau_{BH}$, so that  
\begin{eqnarray}
    q &=& K \tau\sim \frac{M-\frac{Q^2}{r}}{r^3} \sqrt{1 - \frac{2M}{r} + \frac{Q^2}{r^2}}\ \tau_{BH}, 
    \label{qgr}
\end{eqnarray}
which has a minimum in $r$ just outside the outer horizon.   The minimum sets the radius of $P$, while $q\sim1$ at this minimum sets $\tau_{BH}$ \cite{Haggard2014}.
\end{enumerate}

One way or the other, at some point $P$ outside the horizon the dynamics of the gravitational field enters the quantum tunnelling region.  By causality, this region must then be in the causal future of $P$.  In the future of this region, we can assume spacetime to be described again by a solution of the Einstein equations.  Depending on the interpretation of quantum theory one prefers, this can be seen as a de-cohered many world branch, or the result of a (``measuring") interaction with other degrees of freedom (the ``observer") after the quantum process. 

Notice that the entire time-like singularity and the non-causal region of the Reissner-Nordstr\"{o}m spacetime past the  Cauchy horizon (the boundary of the causal future of the singularity) disappear, replaced by the quantum region: a part from the quantum transition, the spacetime is regular. 

This completes the construction of the Reissner-Nordstr\"{o}m black to white hole transition spacetime. 

Before concluding this section, we  observe that the above result can also be taken as a possible form of the effective metric in the {\bf C} and {\bf A} regions in the case of \textit{vanishing} electrical charge. (Another simple guess for the effective metric in the $A$ region can be obtained by replacing $r$ with $\sqrt{r^2+L^2}$ in the Schwarzschild metric \cite{DAmbrosio2018}: see also \cite{Franzin2021} for the same idea).
Loop quantum cosmology \cite{Agullo2013a} suggests that the dominant quantum effect at high curvature is a repulsive force, precisely as in the Reissner-Nordstr\"om geometry (where repulsion does not act on charged matter only). This is the short scale quantum pressure which is also responsible for the quantum bounce of a Planck star \cite{Rovelli2014ps,Rovelli2017f}. Hence the Reissner-Nordstr\"om geometry can also be taken as a guess for the quantum corrections to the metric at short radius, for an effective value of a ``charge" $Q_{eff}^2$ determined by $M$ (and the Planck mass $M_{Planck}$). This can be estimated assuming that the corrections becomes relevant when the curvature is Planckian, which gives 
\begin{equation}
    Q_{eff}\sim M^{\frac23}.
\end{equation}
Hence, restoring physical units
\begin{equation}
    \eta_{eff}\sim \left(\frac{M}{M_{Planck}}\right)^{-\frac13}.
\end{equation}
For a macroscopic mass $M$, $\eta_{eff}\ll1$ and so the resulting metric is very close to Schwarzschild, but since, as pointed out above,  the $Q\to 0$ and $r\to 0$ limits do not commute, the global structure of the metric is radically changed nevertheless. 
If this is correct, the actual effect of quantum gravity makes the Reissner-Nordstr\"om geometry studied here more realistic than the Schwarzschid geometry. 

\section{Cauchy horizon instabilities}
\label{insta}

There are three important physical phenomena that the model defined in this paper disregards: rotation, instabilities and the Hawking radiation. In this last part of the paper we briefly comment on their possible effect.

To account for rotation, we must study the Kerr-Newman metric. As already noticed, the Carter-Penrose maximal extension of the  Kerr-Newman metric is similar to the one of the maximal extension of the Reissner-Nordstr\"{o}m metric. It is therefore reasonable to expect that some qualitative aspects of the black to white bounce extend to the rotating case.   This will be done elsewhere. 

Instabilities are expected before the Cauchy horizon \cite{simpson1973internal,dafermos2003stability}. There is a simple argument that illustrates why.  Consider a sequence of pulses emitted radially at regular time intervals from past infinity. They all reach an observer moving towards the Cauchy horizon in a finite proper time.  Thus, if perturbations enter the black hole for an arbitrary long external time, they all pile up in the finite time of the observer before the  Cauchy horizon. In turn, this is likely to cause a concentration of energy near the Cauchy horizon.  This is going to generate a strong curvature, which at some point may become Planckian.  

Poisson and Israel \cite{poisson1989inner, poisson1990internal} have in fact shown that small perturbations of the metric outside the black hole during or after the star collapsing become unstable once they penetrate \textit{inside} the black hole and lead the Cauchy horizon to become a null curvature singularity. This phenomenon is known as \textit{mass inflation} and is confirmed by numerical investigations~\cite{brady1995black, brady1999internal, burko1997structure, burko1998analytic, Carballo-Rubio2021, Carballo-Rubio2018a}.  Furthermore, physical and numerical indications point to the expectation that the perturbation grows into a \emph{spacelike} singularity in the classical theory. 

Before diverging the curvature must become Planckian. Hence, the region along the Cauchy horizon is situated inside a quantum gravity region. This result  strengthen the hypothesis that the spacetime dynamics enters a quantum region before the development of the Cauchy horizon, as in the hypothesis of this paper, and no Cauchy horizon develops in reality.   Quantum gravity should correct not only the singularities but also the lack of global hyperbolicity of classical general relativity (see also \cite{Carballo-Rubio2021, Carballo-Rubio2018a}.). 

A quick estimate of where the instability drives the spacetime dynamics into the quantum region can be obtained as follows.  Near the inner horizon, the Reissner-Nordstr\"{o}m metric can be written as   (see \cite{poisson2002advanced}) 
\begin{equation}
ds^2 \simeq - 2 e^{-\kappa_- v} e^{\kappa_- u}du dv + r^2 d\Omega^2
\end{equation}
where $\kappa_-=\frac12 f'(r_-)$ is the surface gravity on the inner horizon. The proper time interval $d\tau$ between the reception of signals by  an observer approaching the Cauchy horizon along a (timelike) constant $u+v=2t$ trajectory inside the black hole is related to the proper time interval $dv$ of emission of these signals from infinity by
\begin{equation}
    d\tau \simeq \sqrt{2}\ e^{\kappa_- t} e^{-\kappa_- v} dv 
\end{equation}
approaching the Cauchy horizon ($v\to\infty$), giving a blue shift growing exponentially as 
\begin{equation}
    \frac{d\tau}{dv} \sim e^{-\kappa_- v}
    \label{blueshift}
\end{equation}
This blue shift is expected to hold whatever is the observer motion across the inner horizon~\cite{brady1999internal}.  According to \cite{brady1995black, brady1999internal}, the perturbation generated by this instability gives rise, for $v \to \infty$, to a curvature of the order of  
\begin{equation}
K^2 = g(u) \frac{e^{2 \kappa_- v}}{v^{2q}}
    \label{curvatureinstability}
\end{equation}
where $g(u)$ is a function that vanishes at the outer horizon and increases monotonically and $q$ is a positive integer depending on the characteristics of the in-falling perturbations. Since $v$ is large, the term $v^{2q}$ can be neglected and we have 
\begin{equation}
    v \simeq -\frac{\log {g(u)}}{2\kappa_-} 
    \label{boundaryQGR}
\end{equation}
So we may expect that the spacetime enters the quantum region just because of the instabilities. Since this formula is only reliable in the large $v$ limit, is not clear to us whether this can happen before the onset of the quantum regime studied above. 

But notice that in any case these instabilities happens only in the interior of the black hole as $g(u)$ vanishes on the outer boundary. The black to white transition requires that the region affected by quantum gravity  leaks outside the horizon.  Without this, the apparent horizon becomes an event horizon and the black hole remains such forever, if observed from the exterior. 
 
As Reissner-Nordstr\"{o}m and Kerr share the same causal diagram, the existence of the infinite blue shift and the consequences discussed above, i.e the instability the (Kerr) metric, the null weak singularity along the Cauchy horizon, and its shrinking up to a spacelike singularity at $r = 0$, are also present for the rotating black hole and the charged and rotating Kerr-Newman black hole. 

\begin{figure}
    \centering
    \includegraphics[width=.4\linewidth]{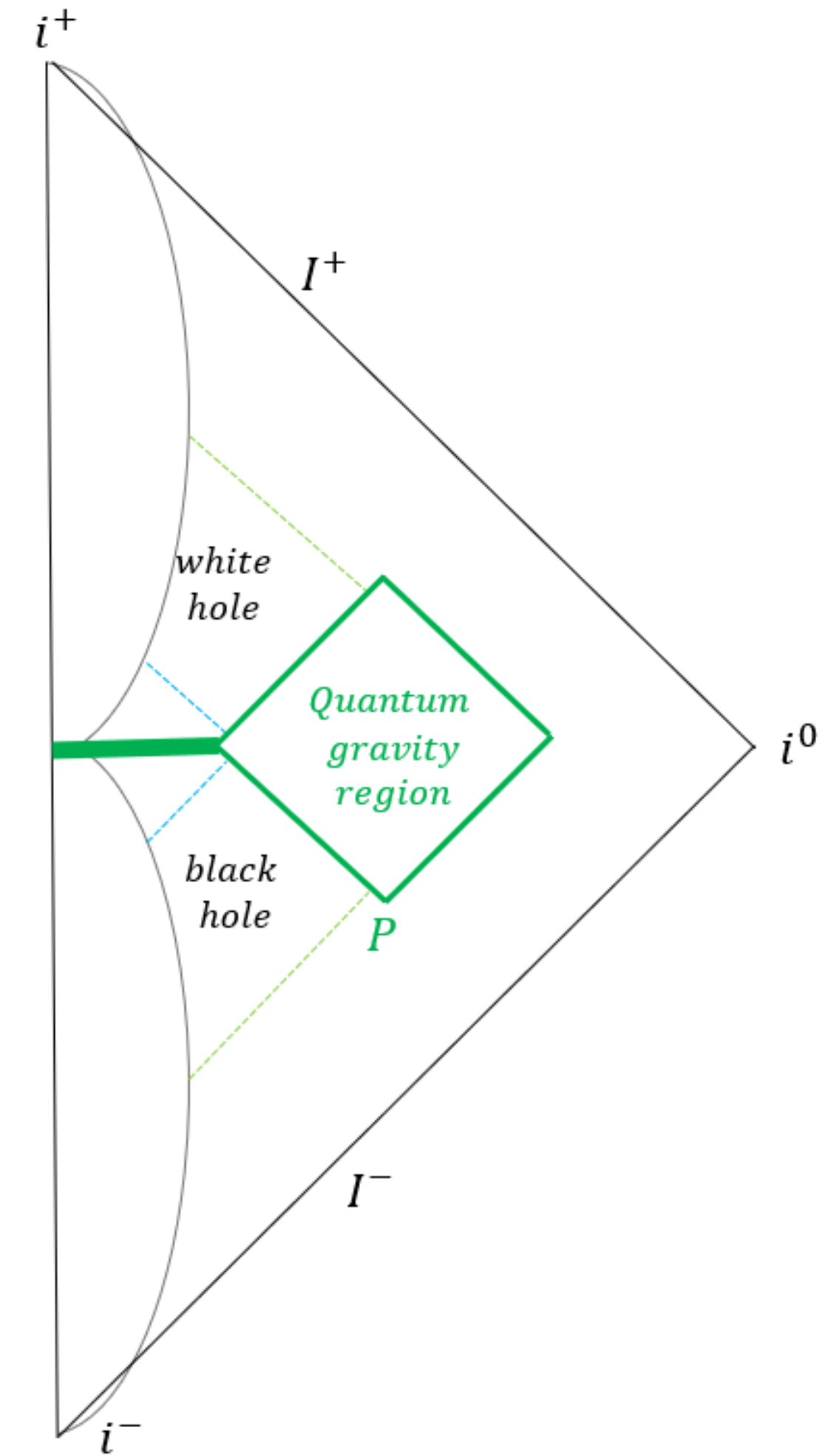}
    \caption{The Carter-Penrose diagram for the black to white transition of the charged black hole when the the quantum gravity region (in green) extends up to the entire internal region. The  difference with Figure~\ref{full} is that here the star encounters a quantum gravity region, and  the white hole region is not classically connected to the black hole region anymore.}
    \label{glueSch}
\end{figure}

Notice that there is also the possibility that the quantum gravity region generated by the instabilities extend all the way to the trajectory of the bouncing star, in which case the physics would be more similar to the one studied in~\cite{Haggard2014}.  See Figure \ref{glueSch}.

\section{Hawking evaporation}
\label{HR}

In the Schwarzschild case, the mass of the hole decreases because of the Hawking Radiation (HR) falling into the hole.  What about its charge ? 

One may speculate that the HR is dominated by massless particles that can escape to infinity more easily. Massless particles in the standard model are not charged, so they may not carry  charge away and diminish the black hole charge $Q$.  The same is not true for the rotating case, because photons can carry angular momentum, although  the HR might still be dominated by $s$-waves. According to \cite{hiscock1990evolution} charged black holes do not necessarily evolve towards the Schwarzschild limit contrary to Kerr black holes. If they are massive enough and if their charge is large enough ($\frac{3}{4} < \eta^2 < 1$ according to~\cite{hiscock1990evolution}), they evolve toward the extremal limit ($\eta$ = 1).  If so, the ratio $\eta=Q/M$ increases at the end of the Hawking evaporation, making the Reissner-Nordsrt\"om model more realistic than the Schwarzschild model for the transition. 

In the absence of Hawking evaporation, the black to white transition might take exponentially long times to happen.  The Hawking evaporation shrinks the mass, making it increasingly more probable, hence likely to happen within a time $M_0^3$. But part of the HR falls into the black hole, affecting the internal metric.  Some consequences of the backreaction of the  HR on the bounce have been explored in \cite{Martin-Dussaud2019}. In the rest of this section we present some sketchy considerations on the possible effect of this back-reaction in the charged case.  These are speculative, because not much is solidly known about the energy-momentum tensor of the HR falling inside the hole and its back-reaction.  For simplicity, we assume here that the ratio $\eta = \frac{Q}{M}$ remains constant. If so, both the outer $r_+ = M(1 + \sqrt{1 - \eta^2})$ and inner $r_- = M(1 - \sqrt{1 - \eta^2})$ horizons decrease.

\begin{figure}
    \centering
    \includegraphics[width=0.30\linewidth]{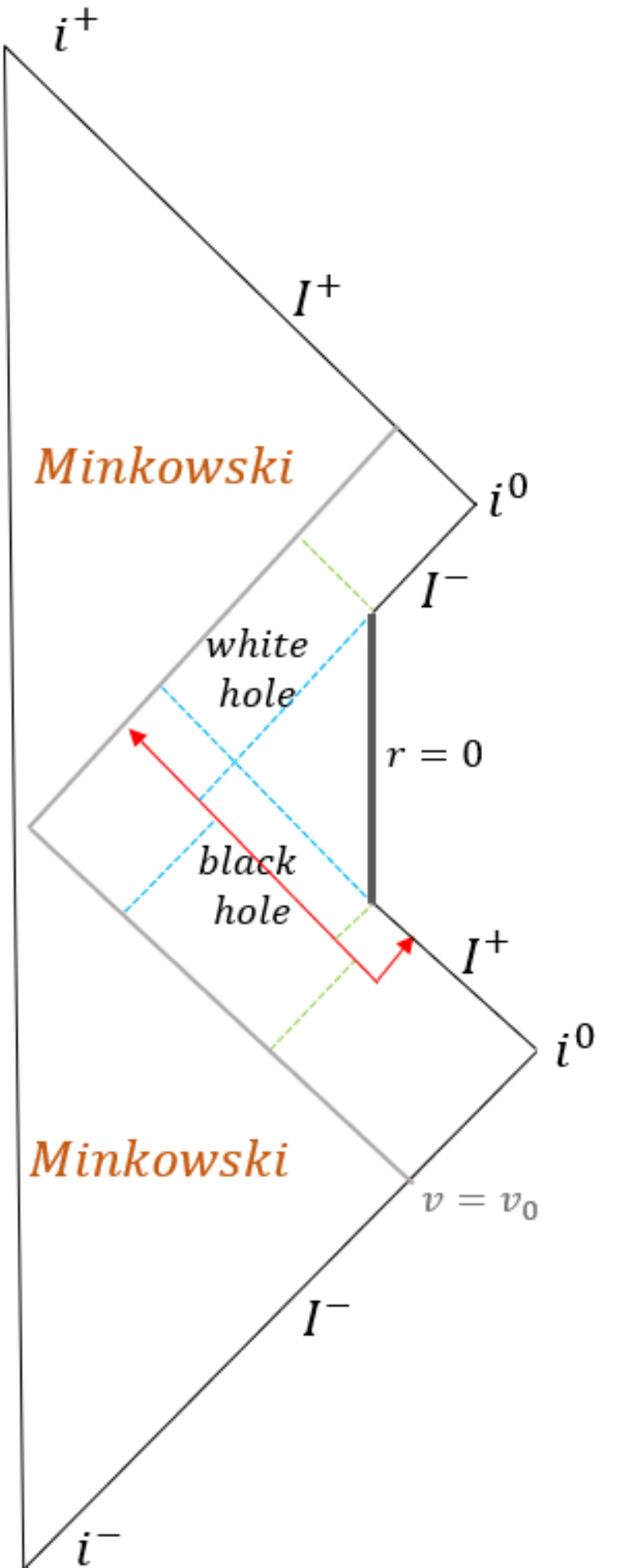}
    \caption{A pair of particles is created just outside the horizon (red lines). One, with mass $+\delta m$ and charge $+\delta q$ escapes to null infinity. The other, with mass $-\delta m$ and charge $-\delta q$ falls into the black hole along a constant $v$ null line. The infalling particle modifies the metric inside the black hole in its future, producing a Reissner-Nordstr\"{o}m metric of a black hole of mass $M - \delta m$ and charge $Q - \delta q$. Both inner and outer horizons shrink. (Here and in the next figure depicted without the gluing of the asymptotic regions.}
    \label{HRcharge}
\end{figure}

There are two simple models of the HR falling inside the hole.  The first is that it is made by negative energy quanta moving along ingoing null geodesics. The second is that it is made by negative energy quanta moving along outgoing null geodesics (the radius of which is still decreasing, since it is a trapped region).  Reality is probably between the two \cite{Hiscock1981,Martin-Dussaud2019,parikh2000hawking}.

Consider an Hawking quantum of mass $-\delta m$ and charge $-\delta q$ in-falling along a constant $v$ null line.  A simple estimate of its backreaction is to modify the geometry by shifting the Reissner-Nordstr\"{o}m mass and charges by these amounts in its future. Notice that this shifts the two horizons inward. Let $r_-'$ be the radius of the new inner horizon.  See Figure \ref{HRcharge}.

Call $c$ the point where the in-falling quantum meets the bouncing star and $r_c$ the radius of this point.  There are  two possibilities.  If $ r_-' > r_c$, the negative energy quanta crosses the new inner horizon $r_-'$ before reaching the star. This does not change the overall picture much. 

\begin{figure}
    \centering
    \includegraphics[width=.36\linewidth]{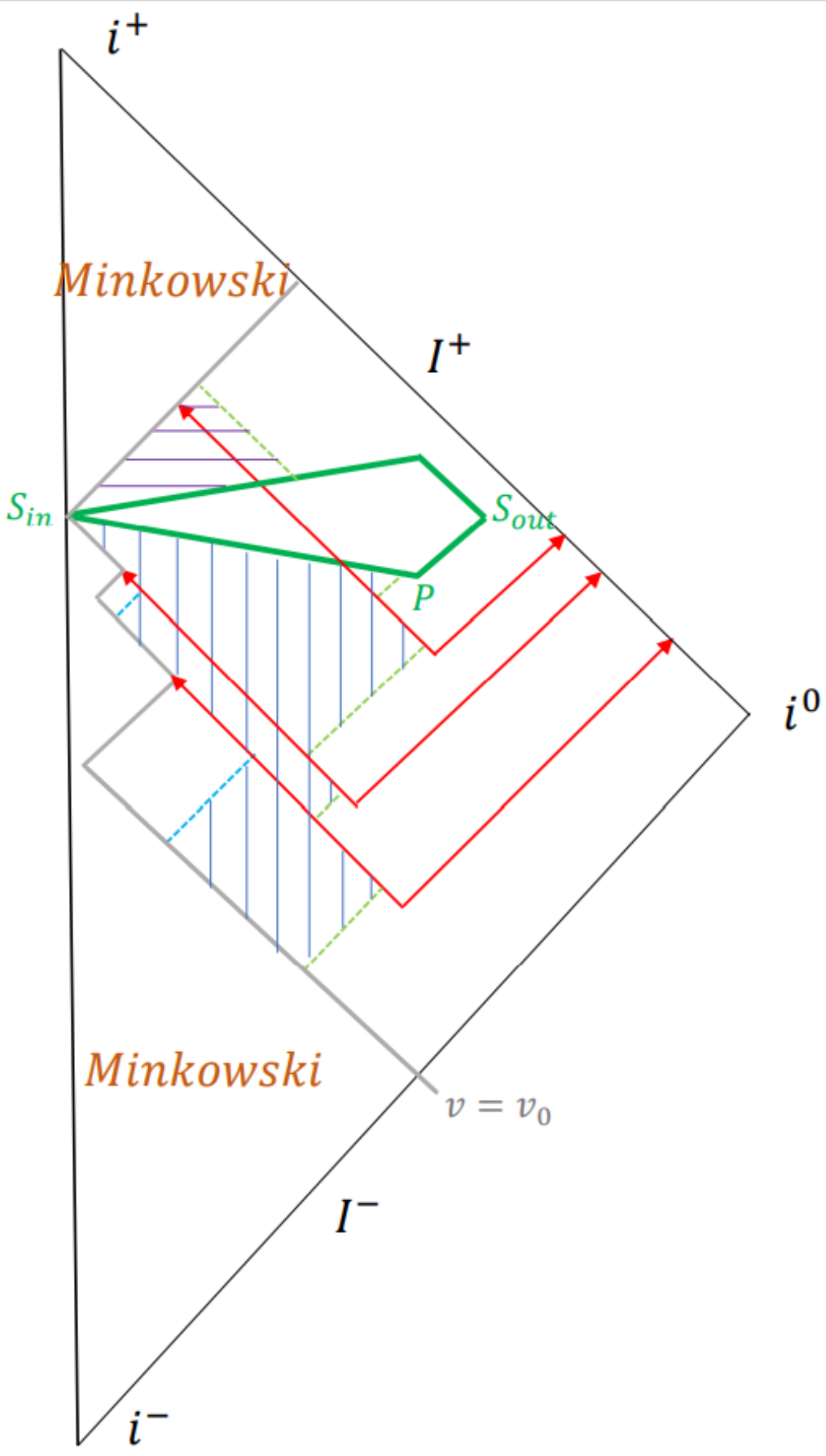}
    \caption{The Hawking radiation shifts the inner horizon for each quanta. A shell (grey line) falls again and bounces at increasingly smaller radius. The vertically hatched area is trapped, the horizontally hatched area  anti-trapped.  Radiation continues to be emitted outside until the point $P$ outside the trapped region becomes planckian}
    \label{HR2B2W}
\end{figure}

However, it might also happen that $ r_-' < r_c$, namely the negative energy in-falling quanta meets the bouncing star \emph{before} entering the new inner horizon. If so, $c$ is in a trapped region hence the star has to fall again, until it crosses new inner horizon and be allowed to bounce at a point $b'$ with radius $r_{b'} < r_b$.  This process might disrupt the classical throat.  In particular, notice that if $r_P$ is near Planckian, so must be the outer and inner horizons at $v_P$, and hence so must be the corresponding $r_c$, which implies that the star itself must have entered a quantum regime.  See Figure \ref{HR2B2W}.  While in the non charged case the infalling HR always encounters the star in the antitrapped region,  in the charged case it is the HR that may make its charge and mass decrease to Planckian values before the final bounce.  
Our control of the energy momentum tensor of the HR and its back-reaction is still insufficient to judge if this is truly the case. 

An argument can be given, suggesting that the onset of the quantum gravity region may be spacelike, especially for small $\eta$ (on this, see also \cite{Bianchi2018d}).  As ingoing HR falls into the hole, it carries negative energy and therefore decreases the local mass of the geometry.

We may expect the metric to be a Reissner-Nordstr\"{o}m metric locally but with a decreasing mass function $M(v)$.  On the inner horizon, the curvature of a given slice (from eq.~\eqref{Kritchmaninvar}) is 
\begin{equation}
    K^2 = 48 \frac{\bigg( 1 - \frac{\eta^2}{1 -\sqrt{1 - \eta^2}}\bigg)^2}{M(v )^4 (1 - \sqrt{1 - \eta^2})^6} \simeq \frac{3072}{M(v)^4 \eta^{12}}
    \label{curvinner}
\end{equation}
where the last approximation holds if $\eta$ is small. This indicates when the inner horizon enters a region of planckian curvature. Let $S_{in}$ be the point where the curvature becomes planckian on the inner horizon and $v_{in}$ its advanced time. This is the innermost point included in the quantum region.   The curvature at the \emph{outer} horizon $r_+ = M(v)(1 +\sqrt{1 - \eta^2})$ is 
\begin{equation}
    K^2 = 48 \frac{\bigg( 1 - \frac{\eta^2}{1 +\sqrt{1 - \eta^2}}\bigg)^2}{M(v)^4 (1 + \sqrt{1 - \eta^2})^6} \simeq \frac{3}{4} \frac{1}{M(v)^4}
    \label{curvouter}
\end{equation}
where again the last approximation holds if $\eta$ is  small. Comparing~\eqref{curvinner} 
and~\eqref{curvouter} for  $K^2 \sim 1$,
\begin{equation}
    \frac{M(v_{in})}{M(v_{P})} = \frac{8}{\eta^3} > 1
\end{equation}
Hence, $v_{{in}} < v_{P}$ as $M$ is a decreasing function of $v$ and so the line joining the points $P$ and $S_{in}$ is spacelike.  The line joining $P$ and $S_{in}$ can be taken as the boundary of the quantum region.  This picture has similarities with the Schwarzschild black to white transition. 

\begin{figure}[t]
    \centering
    \includegraphics[width=0.35\linewidth]{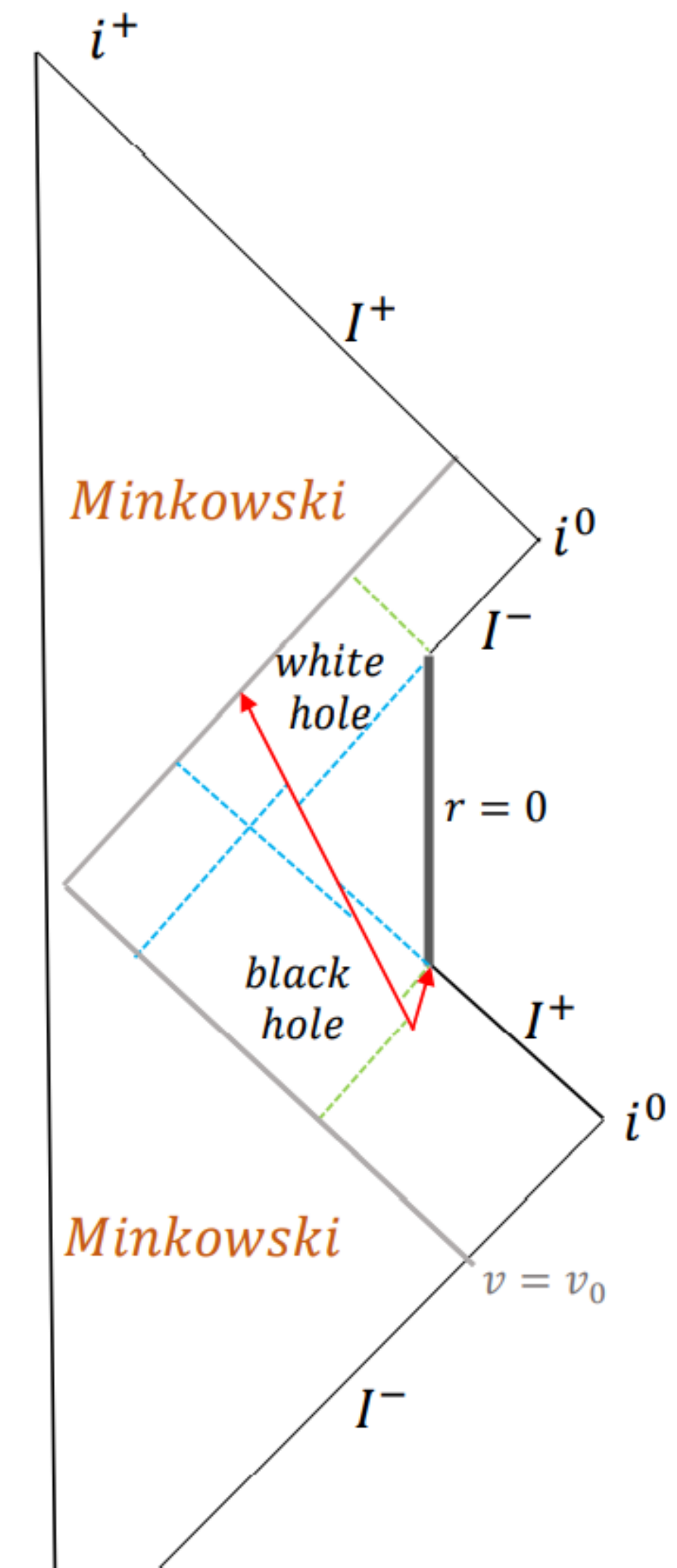}
    \caption{Massive Hawking radiation quanta enter the quantum region, and make the inner (and outer) horizon shrinks. (Depicted without the gluing of the two asymptotic regions.)}
    \label{HR3B2W}
\end{figure}

Finally, outgoing HR quanta inside the hole, on the other hand, fall into the quantum region.  The negative energy they carry can emerge into the white hole and decrease the mass and charge of the star inside the white hole and the classical throat is not destroyed. See \cite{Martin-Dussaud2019,martin2019evaporating}.   The same may happen if the Hawking quanta are massive (see Figure \ref{HR3B2W}.)

\vfill


\section{Conclusion}

The presence of charge renders the black to white transition more interesting than the non charged case.  The bounce of the star and the region immediately surrounding it evolve into a white hole simply by following the classical dynamics.   Only the horizon region  tunnels.   The white hole is in the same location of the same asymptotic region as the black hole that originates it. 

The extension of the spacetime region surrounding the star that evolves into a white hole classically depends on the charge: in the limit of vanishing charge, the situation is similar to the Schwarzschild tunnelling. 

We have also tentatively explored the effects of the classical instabilities and the back-reaction of the HR on the process.  These may alter the picture of the interior, creating a spacelike onset of the quantum region (as in the Schwarzschild case) and decreasing the range of the classical region. But they do not seem to alter the basic possibility of black to white hole quantum tunnelling.  

The qualitative similarities of the Reissner-Nordstr\"{o}m and Kerr-Newman metrics suggest that the entire black to white hole quantum tunnelling may be a general possibility, and therefore represent a likely scenario for the fate of all real black holes.


\bibliographystyle{utcaps}
\bibliography{Reissner-Nordstrom}

\end{document}

%% file: pack_def.tex
\usepackage{amssymb}
\usepackage{amsmath}
\usepackage{amsfonts}
\usepackage[colorlinks,linkcolor=red,urlcolor=blue,citecolor=blue]{hyperref}
\usepackage{graphics}
\usepackage{tikz}  
\usetikzlibrary{arrows,shapes,positioning,shadows,backgrounds,fit}
\usepackage{braket}

%% file: article.bbl
\providecommand{\href}[2]{#2}\begingroup\raggedright\begin{thebibliography}{10}

\bibitem{Haggard2014}
H.~M. Haggard and C.~Rovelli, ``{Quantum-gravity effects outside the horizon
  spark black to white hole tunneling},'' {\em Physical Review D} {\bfseries
  92} no.~10, (2015) 104020, \href{http://arxiv.org/abs/1407.0989}{{\ttfamily
  arXiv:1407.0989}}.

\bibitem{Rovelli2014ps}
C.~Rovelli and F.~Vidotto, ``{Planck stars},'' {\em Int. J. Mod. Phys. D}
  {\bfseries 23} (2014) 1442026,
  \href{http://arxiv.org/abs/1401.6562}{{\ttfamily arXiv:1401.6562}}.

\bibitem{christodoulou2016realistic}
M.~Christodoulou, C.~Rovelli, S.~Speziale, and I.~Vilensky, ``{Realistic
  observable in background-free quantum gravity: the Planck-star
  tunnelling-time},'' {\em Phys. Rev. D} {\bfseries 94} (2016) 084035,
  \href{http://arxiv.org/abs/1605.05268}{{\ttfamily arXiv:1605.05268}}.

\bibitem{Rovelli2018h}
C.~Rovelli, ``{Black Hole Evolution Traced Out with Loop Quantum Gravity},''
  {\em Physics} {\bfseries 11} (Jan, 2018) 127,
  \href{http://arxiv.org/abs/1901.04732}{{\ttfamily arXiv:1901.04732}}.

\bibitem{Bianchi2018e}
E.~Bianchi, M.~Christodoulou, F.~D'Ambrosio, H.~M. Haggard, and C.~Rovelli,
  ``{White holes as remnants: A surprising scenario for the end of a black
  hole},'' {\em Classical and Quantum Gravity} {\bfseries 35} no.~22, (2018) ,
  \href{http://arxiv.org/abs/1802.04264}{{\ttfamily arXiv:1802.04264}}.

\bibitem{Christodoulou2018d}
M.~Christodoulou and F.~D'Ambrosio, ``{Characteristic time scales for the
  geometry transition of a black hole to a white hole from spinfoams},''
  \href{http://arxiv.org/abs/1801.03027}{{\ttfamily arXiv:1801.03027}}.

\bibitem{Rovelli2018a}
P.~Martin-Dussaud and C.~Rovelli, ``{Interior metric and ray-tracing map in the
  firework black-to-white hole transition},'' {\em Classical and Quantum
  Gravity} {\bfseries 35} no.~14, (2018) ,
  \href{http://arxiv.org/abs/1803.06330}{{\ttfamily arXiv:1803.06330}}.

\bibitem{Martin-Dussaud2019}
P.~Martin-Dussaud and C.~Rovelli, ``{Evaporating black-to-white hole},'' {\em
  Classical and Quantum Gravity} {\bfseries 36} no.~24, (May, 2019) ,
  \href{http://arxiv.org/abs/1905.07251}{{\ttfamily arXiv:1905.07251}}.

\bibitem{Dambrosio2021}
F.~D'Ambrosio, M.~Christodoulou, P.~Martin-Dussaud, C.~Rovelli, and F.~Soltani,
  ``{The End of a Black Hole's Evaporation - Part I},'' {\em Phys. Rev. D}
  {\bfseries 103} (2021) 106014,
  \href{http://arxiv.org/abs/2009.05016}{{\ttfamily arXiv:2009.05016}}.

\bibitem{soltani2021end}
F.~Soltani, C.~Rovelli, and P.~Martin-Dussaud, ``{The End of a Black Hole's
  Evaporation - Part II},'' {\em Phys. Rev. D} {\bfseries to appear} (2021) ,
  \href{http://arxiv.org/abs/2105.06876}{{\ttfamily arXiv:2105.06876}}.

\bibitem{Barrau2014b}
A.~Barrau, C.~Rovelli, and F.~Vidotto, ``{Fast radio bursts and white hole
  signals},'' {\em Physical Review D} {\bfseries 90} no.~12, (2014) 127503,
  \href{http://arxiv.org/abs/1409.4031}{{\ttfamily arXiv:1409.4031}}.

\bibitem{Barrau2014c}
A.~Barrau and C.~Rovelli, ``{Planck star phenomenology},'' {\em Physics Letters
  B} {\bfseries 739} (2014) 405--409,
  \href{http://arxiv.org/abs/1404.5821}{{\ttfamily arXiv:1404.5821}}.

\bibitem{Vidotto2016}
F.~Vidotto, A.~Barrau, B.~Bolliet, M.~Shutten, and C.~Weimer,
  ``{Quantum-gravity phenomenology with primordial black holes},''
  \href{http://arxiv.org/abs/1609.02159}{{\ttfamily arXiv:1609.02159}}.

\bibitem{Rovelli2017f}
C.~Rovelli, ``{Planck stars: new sources in radio and gamma astronomy?},'' {\em
  Nature Astronomy} no.~1, (2017) 0065,
  \href{http://arxiv.org/abs/1708.01789}{{\ttfamily arXiv:1708.01789}}.

\bibitem{Barrau2017}
A.~Barrau, B.~Bolliet, M.~Schutten, and F.~Vidotto, ``{Bouncing black holes in
  quantum gravity and the Fermi gamma-ray excess},'' {\em Physics Letters B}
  {\bfseries 772} (2017) 58--62,
  \href{http://arxiv.org/abs/1606.08031}{{\ttfamily arXiv:1606.08031}}.

\bibitem{Raccanelli2017}
A.~Raccanelli, F.~Vidotto, and L.~Verde, ``{Effects of primordial black holes
  quantum gravity decay on galaxy clustering},'' {\em Journal of Cosmology and
  Astroparticle Physics} {\bfseries 2018} no.~8, (Aug, 2018) ,
  \href{http://arxiv.org/abs/1708.02588}{{\ttfamily arXiv:1708.02588}}.

\bibitem{Vidotto2018a}
F.~Vidotto, ``{Quantum insights on Primordial Black Holes as Dark Matter},'' in
  {\em Proceedings of Science}, vol.~335.
\newblock Sissa Medialab Srl, Nov, 2018.
\newblock \href{http://arxiv.org/abs/1811.08007}{{\ttfamily arXiv:1811.08007}}.

\bibitem{Rovelli2018f}
C.~Rovelli and F.~Vidotto, ``{Small black/white hole stability and dark
  matter},'' {\em Universe} {\bfseries 4} no.~11, (Nov, 2018) 127,
  \href{http://arxiv.org/abs/1805.03872}{{\ttfamily arXiv:1805.03872}}.

\bibitem{Rovelli2018g}
C.~Rovelli and F.~Vidotto, ``{Pre-big-bang black-hole remnants and past low
  entropy},'' {\em Universe} {\bfseries 4} no.~11, (2018) ,
  \href{http://arxiv.org/abs/1805.03224}{{\ttfamily arXiv:1805.03224}}.

\bibitem{Vidotto2018b}
F.~Vidotto, ``{Measuring the Last Burst of Non-singular Black Holes},'' {\em
  Foundations of Physics} {\bfseries 48} no.~10, (Mar, 2018) 1380--1392,
  \href{http://arxiv.org/abs/1803.02755}{{\ttfamily arXiv:1803.02755}}.

\bibitem{Barausse2020}
E.~Barausse and E.~al., ``{Prospects for fundamental physics with LISA},'' {\em
  General Relativity and Gravitation} {\bfseries 52} (2020) ,
  \href{http://arxiv.org/abs/2001.09793}{{\ttfamily arXiv:2001.09793}}.

\bibitem{barrau2021closer}
A.~Barrau, L.~Ferdinand, K.~Martineau, and C.~Renevey, ``{Closer look at white
  hole remnants},'' {\em Physical Review D} {\bfseries 103} no.~4, (Jan, 2021)
  , \href{http://arxiv.org/abs/2101.01949}{{\ttfamily arXiv:2101.01949}}.

\bibitem{Rovelli2017a}
C.~Rovelli, ``{Black holes have more states than those giving the
  Bekenstein-Hawking entropy: a simple argument},''
  \href{http://arxiv.org/abs/1710.00218}{{\ttfamily arXiv:1710.00218}}.

\bibitem{Rovelli2019a}
C.~Rovelli, ``{The subtle unphysical hypothesis of the firewall theorem},''
  {\em Entropy} {\bfseries 21} no.~9, (2019) ,
  \href{http://arxiv.org/abs/1902.03631}{{\ttfamily arXiv:1902.03631}}.

\bibitem{penrose1999question}
R.~Penrose, ``{The question of cosmic censorship},'' {\em Journal of
  Astrophysics and Astronomy} {\bfseries 20} no.~3-4, (1999) 233--248.

\bibitem{poisson1989inner}
E.~Poisson and W.~Israel, ``{Inner-horizon instability and mass inflation in
  black holes},'' {\em Phys. Rev. Lett.} {\bfseries 63} (1898) 1663.

\bibitem{poisson1990internal}
E.~Poisson and W.~Israel, ``{Internal structure of black holes},'' {\em
  Physical Review D} {\bfseries 41} no.~6, (1990) 1796.

\bibitem{christodoulou2018characteristic}
M.~Christodoulou and F.~D'Ambrosio, ``{Characteristic time scales for the
  geometry transition of a black hole to a white hole from spinfoams},'' {\em
  arXiv preprint arXiv:1801.03027} (2018) .

\bibitem{Haggard2016}
H.~M. Haggard and C.~Rovelli, ``{Quantum gravity effects around Sagittarius
  A*},'' {\em International Journal of Modern Physics D} {\bfseries 25} no.~12,
  (2016) , \href{http://arxiv.org/abs/1607.00364}{{\ttfamily
  arXiv:1607.00364}}.

\bibitem{Carballo-Rubio2020}
R.~Carballo-Rubio, F.~D. Filippo, S.~Liberati, and M.~Visser, ``{Geodesically
  complete black holes},'' 2020.

\bibitem{poisson2002advanced}
E.~Poisson, ``{An advanced course in general relativity},'' {\em lecture notes
  at University of Guelph} (2002) .

\bibitem{DAmbrosio2018}
F.~D'Ambrosio and C.~Rovelli, ``{How information crosses Schwarzschild's
  central singularity},'' {\em Classical and Quantum Gravity} {\bfseries 35}
  no.~21, (2018) , \href{http://arxiv.org/abs/1803.05015}{{\ttfamily
  arXiv:1803.05015}}.

\bibitem{Franzin2021}
E.~Franzin, S.~Liberati, J.~Mazza, A.~Simpson, and M.~Visser, ``{Charged
  black-bounce spacetimes},'' {\em Journal of Cosmology and Astroparticle
  Physics} {\bfseries 2021} no.~07, (2021) 036,
  \href{http://arxiv.org/abs/2104.11376}{{\ttfamily arXiv:2104.11376}}.

\bibitem{Agullo2013a}
I.~Agullo and A.~Corichi, ``{Loop Quantum Cosmology},''
  \href{http://arxiv.org/abs/1302.3833}{{\ttfamily arXiv:1302.3833}}.

\bibitem{simpson1973internal}
M.~Simpson and R.~Penrose, ``{Internal instability in a
  Reissner-Nordstr{\"{o}}m black hole},'' {\em International Journal of
  Theoretical Physics} {\bfseries 7} no.~3, (1973) 183--197.

\bibitem{dafermos2003stability}
M.~Dafermos, ``{Stability and instability of the Cauchy horizon for the
  spherically symmetric Einstein-Maxwell-scalar field equations},'' {\em Annals
  of mathematics} (2003) 875--928.

\bibitem{brady1995black}
P.~R. Brady and J.~D. Smith, ``{Black hole singularities: a numerical
  approach},'' {\em Physical review letters} {\bfseries 75} no.~7, (1995) 1256.

\bibitem{brady1999internal}
P.~R. Brady, ``{The internal structure of black holes},'' {\em Progress of
  Theoretical Physics Supplement} {\bfseries 136} (1999) 29--44.

\bibitem{burko1997structure}
L.~M. Burko, ``{Structure of the black hole's Cauchy-horizon singularity},''
  {\em Physical review letters} {\bfseries 79} no.~25, (1997) 4958.

\bibitem{burko1998analytic}
L.~M. Burko and A.~Ori, ``{Analytic study of the null singularity inside
  spherical charged black holes},'' {\em Physical Review D} {\bfseries 57}
  no.~12, (1998) R7084.

\bibitem{Carballo-Rubio2021}
R.~Carballo-Rubio, F.~{Di Filippo}, S.~Liberati, C.~Pacilio, and M.~Visser,
  ``{Inner horizon instability and the unstable cores of regular black
  holes},'' {\em Journal of High Energy Physics} {\bfseries 2021} no.~5, (2021)
  , \href{http://arxiv.org/abs/2101.05006}{{\ttfamily arXiv:2101.05006}}.

\bibitem{Carballo-Rubio2018a}
R.~Carballo-Rubio, F.~{Di Filippo}, S.~Liberati, C.~Pacilio, and M.~Visser,
  ``{On the viability of regular black holes},'' {\em Journal of High Energy
  Physics} {\bfseries 2018} no.~7, (2018) ,
  \href{http://arxiv.org/abs/1805.02675}{{\ttfamily arXiv:1805.02675}}.

\bibitem{hiscock1990evolution}
W.~A. Hiscock and L.~D. Weems, ``{Evolution of charged evaporating black
  holes},'' {\em Physical Review D} {\bfseries 41} no.~4, (1990) 1142.

\bibitem{Hiscock1981}
W.~Hiscock, ``{Models of evaporating black holes. I},'' {\em Phys. Rev. D}
  {\bfseries 23} (1981) 2813----2822.

\bibitem{parikh2000hawking}
M.~K. Parikh and F.~Wilczek, ``{Hawking radiation as tunneling},'' {\em
  Physical Review Letters} {\bfseries 85} no.~24, (2000) 5042.

\bibitem{Bianchi2018d}
E.~Bianchi and H.~M. Haggard, ``{Spin fluctuations and black hole
  singularities: The onset of quantum gravity is spacelike},'' {\em New Journal
  of Physics} {\bfseries 20} no.~10, (2018) ,
  \href{http://arxiv.org/abs/1803.10858}{{\ttfamily arXiv:1803.10858}}.

\bibitem{martin2019evaporating}
P.~Martin-Dussaud and C.~Rovelli, ``{Evaporating black-to-white hole},'' {\em
  Class. Quantum Grav.} {\bfseries 36} (2019) 245002,
  \href{http://arxiv.org/abs/1905.07251}{{\ttfamily arXiv:1905.07251}}.

\end{thebibliography}\endgroup
